\documentclass[twocolumn,prb]{revtex4}  
\usepackage{graphicx}
\usepackage{dcolumn}
\usepackage{amssymb}
\usepackage[T1]{fontenc}
\usepackage{mathptmx}
\usepackage{txfonts}
\usepackage{mathrsfs}
\usepackage{bm}

\begin{document}
\title{Simple derivation from postulates of generalized vacuum Maxwell equations}
\author{Chun Wa Wong}
\email{cwong@physics.ucla.edu}   
\affiliation{Department of Physics and Astronomy, 
University of California, Los Angeles, CA 90095-1547, USA}
\date{\today}

\begin{abstract}
The two postulates of special relativity plus the postulates of conserved 
charges, both electric and magnetic, and a resulting linear system are 
sufficient for the derivation of the generalized vacuum Maxwell equations with 
both charges. The derivative admits another set of Maxwell equations for charges 
that are the opposite-parity partners of the usual electric and magnetic charges. 
These new charges and their photons are parts of the parallel universe of dark 
matter.
\end{abstract}
\maketitle

\section{Introduction }

The purpose of this paper is to display the conceptual basis of classical electromagnetism 
by deriving the usual generalized vacuum (gv) Maxwell equations with both electric and 
magnetic charges from the following four postulates:  
\begin{enumerate}
\item[P1.] {\it Principle of relativity}:  \\
Physical laws are the same in all inertial frames.
\item[P2.] {\it Constancy of light speed}: \\
Light propagates in vacuum with the same speed $c$ in all inertial frames.
\item[P3.] {\it Charge conservation}:\\
Electric and other charges have the same values in all inertial frames. 
\item[P4.] {\it Linearity}:\\
Classical electromagnetic (EM) theory is described by differential equations linear in 
the EM fields.   
\end{enumerate}

P1 and P2 are Einstein's postulates for Special Relativity (SR) needed to ensure that 
the resulting dynamical equations have the same form in all inertial frames. P3 defines 
the essential attribute of the source charges of the EM fields. P4 defines classical EM 
theory as a linear approximation to more complicated physics. Finally, Maxwell equations 
have both retarded and advanced wave solutions. So causality is {\it not} one of the 
required postulates. By not using causality, the present derivation differs significantly
from that given by Heras \cite{Heras09}, although charge conservation is required in both 
derivations, as it should.

The derivation given in Section 2 is simple and elementary and is readily accessible to
readers familiar with the basic properties of 4-vectors in spacetime.

In Section 3, the space-time symmetries of the 4-currents and EM fields in the gvMaxwell 
equations are described. It then becomes clear that the given derivation admits another  
set of gvMaxwell equations that are the opposite-parity partners of the first set. The 
new equations involve opposite-parity charges and photons that are parts of the parallel 
universe of dark matter.

\section{Deriving the usual gvMaxwell equations } 

The present derivation is based on the following idea. The derivation has
to be relativistic to comply with the two SR postulates. In the very intuitive 
Minkowski notation \cite{Sakurai67}, all vectors and tensors carry only 
subscripts with time treated as a space-like component, but made imaginary to 
display its unique role:
\begin{eqnarray}
x &=& ({\bf x},\, x_4 = ict), \nonumber \\
\partial &=& \frac{\partial}{\partial x_\mu}
 = \left(\bm{\nabla},\, \partial_4 = \frac{\partial}{ic\partial t}\right).
\label{Minkowski} 
\end{eqnarray} 
The two inhomogeneous Maxwell equations to be derived can then be written as 
one equation
\begin{eqnarray}
s_GJ_\mu = \partial_\nu F_{\mu\nu}(x), \quad s_G = 4\pi/c,
\label{inhomoEq}
\end{eqnarray} 
where $J_\mu = ({\bf J}, ic\rho)$ is the electric 4-current and 
$F_{\mu\nu}(x) = -F_{\nu\mu}(x)$ is the antisymmetric rank-2 Maxwell field 
tensor containing the EM fields {\bf E} and {\bf B} as their six nontrivial 
elements that can be nonzero. The factor $1/c$ in $s_G$ identifies the 
dimension of $F$ as that of a dipole moment of the charge density $\rho$, 
namely charge per unit area. The appearance of $4\pi$, the total solid angle 
in 3D space, shows that we are using Gaussian (G) units.

Differentiation of Eq. (\ref{inhomoEq}) gives  
\begin{eqnarray}
s_G\partial_\mu J_\mu &=& \partial_\mu\partial_\nu F_{\mu\nu} 
= - \partial_\mu\partial_\nu F_{\nu\mu} \nonumber \nonumber \\
&=& 0.
\label{ConsJ}
\end{eqnarray} 
This is the (Lorentz-) invariant continuity equations that extends electric
charge conservation ($d\rho/dt = 0$) for the charge density $\rho$ in the laboratory 
frame to all inertial frames. Thus Maxwell equations imply charge conservation,
a well-known result obtained in many textbooks. See, for example, Sakurai 
\cite{Sakurai67-2}. 

In deriving the gvMaxwell equations from postulates, we need to invert the process by 
going from Eq. (\ref{ConsJ}) to Eq. (\ref{inhomoEq}). There are logical gaps big and 
small that have to be bridged. We shall proceed in two steps: (i) Derive the 
inhomogeneous (electric source) equation \cite{Kapuscik09}, and (ii) derive 
the inhomogeneous (magnetic source) equation when magnetic sources are also present. 
The standard Maxwell equations without magnetic sources are then obtained as the 
special case of zero magnetic 4-current.

At the risk of some repetition, let us start from the beginning. In one 
inertial frame, charge conservation can be expressed conveniently in 
the differential form
\begin{eqnarray}
\frac{d\rho}{dt} = 0 
= \partial_t \rho  + \bm{\nabla \cdot}{\bf J}, 
\quad \partial_t = \frac{\partial}{\partial t}, \quad {\bf J} \equiv {\bf v}\rho.
\label{continuity}
\end{eqnarray} 
This continuity equation becomes explicitly frame-independent when written in 
the invariant form $\partial_\mu J_\mu(x) = 0$.

Any conserved 4-current $J_\mu$ has the structure of Eq. (\ref{inhomoEq}),
$s_GJ_\mu = \partial_\nu F_{\mu\nu}(x)$, where $F_{\mu\nu}(x) = -F_{\nu\mu}(x)$ 
is an antisymmetric rank-2 tensor. This is because Eq. (\ref{ConsJ}) is then 
guaranteed. 

A rank-2 spacetime tensor can be displayed as a 4D matrix. If the matrix $F$ is 
antisymmetric, it has zero diagonal elements and only 6 independent off-diagonal 
elements that can be nonzero. Eq. (\ref{ConsJ}) thus contains 12 terms in 6 
canceling pairs.

A 4D spacetime matrix contains a 3D spatial submatrix at its ``core'', bordered by 
an extra row and column involving time. In the Minkowswki notation, the 2-sided 
border contains the last row and the last column. It is convenient to make all 
purely spatial quantities real in physics, because many of them are measurable 
experimentally in terms of real numbers. So the antisymmetric field 
tensor/matrix $F$ (or $F_{\mu\nu}$ in the so-called index notation) can be 
defined to have a real spatial part $F_{j\,k}, \; j,\, k =1,2,3$. This means 
that a border element is imaginary in the Minkowski notation, because it carries 
only one time index. It is timelike if the spatial elements in the 3D core
are spacelike. All elements of matrix $F$ have the same dimension.

The three nontrivial elements residing in the border row defines a 3D vector 
\begin{eqnarray}
F_{4j} = iE_j,
\label{Evector}
\end{eqnarray} 
where ${\bf E}$ has been taken to be real. By antisymmetry, the border column 
elements are $F_{j\,4} = -iE_j$. 

The remaining 3 nontrivial, off-diagonal elements reside in the spatial core,
and are antisymmetric in two spatial indices $j\,k$. They too can be 
made into a 3D vector by using the 3D permutation (Levi-Civita) symbol:
\begin{eqnarray}
F_{j\, k} = \varepsilon_{j\,k\ell}B_\ell.
\label{Bvector}
\end{eqnarray}
This spatial antisymmetry gives rise to a rotation that is responsible for the 
Lorentz force and the Hall effect. {\bf B} is in general different from {\bf E},
as we shall explain in Section \ref{symmetries}.  

We now apply the linearity postulate P4 by identifying the vectors {\bf E} and 
{\bf B} as the EM fields themselves. It is useful to display separately the time 
and space parts of the dynamical Eq. (\ref{inhomoEq}):
\begin{eqnarray}
s_G J_4 &=& \partial_j F_{4j}(x), \nonumber \\
s_G J_j &=& \partial_\nu F_{j\,\nu}(x).
\label{J-rt}
\end{eqnarray} 
In terms of the two 3-vectors ${\bf E}$ and ${\bf B}$ contained in $F$, these 
equations read
\begin{eqnarray}
4\pi \rho(x) &=& \bm{\nabla \cdot} {\bf E}(x), \nonumber \\
4\pi J_j &=& c(\bm{\nabla \times} {\bf B})_j  - \partial_t E_j.
\label{inhomoEq-rt}
\end{eqnarray} 
These are {\it linear}, inhomogeneous dynamical equations for the EM fields driven by an
arbitrary conserved current $J_\mu$. For electric charges in particular, they state the 
Gauss and Amp\`ere--Maxwell EM laws (summaries of experimental facts), respectively. 

Thus charge conservation implies the two inhomogeneous dynamical (Maxwell) 
equations if the EM fields {\bf E} and {\bf B} appear only linearly. These are 
the dynamical equations when the field tensor $F$ has the structure
\{{\bf E}, {\bf B}\} specified by the vectors appearing in the border and
in the spatial core, respectively. 

Are there other solutions? First note that it is the border vector, here {\bf E}, that 
betrays the presence of the source electric charge density $\rho$. Since $\rho$ can 
contain both positive and negative charges, {\bf E} can always be used without an 
extra negative sign. This leaves only one other tensor structure \{{\bf E}, 
-{\bf B}\} for the electric 4-current. The resulting solution is mathematically 
and physically the same as the first solution but uses a left-handed convention 
for all rotations (curls and cross products). Hence the first solution for 
electric 4-currents is physically unique when {\bf E} is in the border.

However, $F$ is not the only EM field tensor that can be constructed from the given EM 
fields. The second and only remaining possibility is to relocate {\bf B} to the border 
and {\bf E} to the core. With {\bf B} in the border, we are now concerned with magnetic 
4-currents driving the {\it same} EM fields. Two sign choices are possible for the
core vector: $\{{\bf E}_m = {\bf B}, {\bf B}_m = \pm{\bf E}\}$. They differ in their 
Poynting vectors giving the directions of their EM waves and energy transports:
\begin{eqnarray}
s_G{\bf S}_m &=& {\bf E}_m \bm{\times} {\bf B}_m 
= \pm {\bf B} \bm{\times} {\bf E} \nonumber \\
&=& \mp s_G{\bf S}_e
\label{Poytings}
\end{eqnarray} 
The solution with waves moving in the same direction as waves generated by 
the partner electric 4-current comes from the tensor structure 
$\{{\bf B}, -{\bf E}\}$. The fields and currents of this {\it magnetic} solution, 
called the {\it dual} partner of the electric solution, are marked by left 
subscripts $_*$. The {\it duality transformation} of the original EM fields that 
makes explicit the magnetic 4-current contribution is then:
\begin{eqnarray}
\{ {\bf E}, {\bf B}\} & \rightarrow & 
\{ _*{\bf E}, \,_*{\bf B}\} \equiv \{ {\bf B}, -{\bf E}\}, \nonumber \\
J & \rightarrow & \; _*J, 
\label{*Transf}
\end{eqnarray} 
where both currents can be chosen arbitrarily. The resulting dual dynamical 
equation is just
\begin{eqnarray}
s_G\, _*J_\mu = \partial_\nu \,_*F_{\mu\nu}(x).
\label{inhomoEq*}
\end{eqnarray} 
Its time and space parts can be read off from Eq. (\ref{inhomoEq-rt}) with
the duality substitution made on the fly:
\begin{eqnarray}
4\pi \,_*\rho &=& \bm{\nabla \cdot} {\bf B}, \nonumber \\
4\pi \, _*J_j &=& -c(\bm{\nabla \times} {\bf E})_j  - \partial_t B_j.
\label{inhomoEq*-rt}
\end{eqnarray} 
This completes the derivation of the gvMaxwell equations from the given four 
postulates.  

When magnetic charges are absent, Eq. (\ref{inhomoEq*-rt}) gives the standard 
homogeneous equations describing the absence of magnetic charges and the Faraday law, respectively. 

The gvMaxwell equations have an additional duality property not shown in 
Eq. (\ref{*Transf}). On taking its dual, we have
\begin{eqnarray}
\{ _*{\bf E}, \,_*{\bf B}\} & \rightarrow & 
\{ _*{\bf B}, \,-_*{\bf E}\} = \{ -{\bf E}, -{\bf B}\} , \nonumber \\
_*J & \rightarrow & \; _*(_*J) = -J. 
\label{**Transf}
\end{eqnarray} 
The $_*(_*J)= -J$ relation is useful for sign checking when working with gvMaxwell 
equations. The dual $_*J$ is just a Hodge dual \cite{Jackson99, Penrose05} in the 
mathematics of differential forms. Our derivation of the duality transformation is 
completely self-contained, however, and does not require a knowledge of differential 
forms.

What about the first sign choice in Eq. (\ref{Poytings}) for the field tensor with 
the structure $\{{\bf B}, {\bf E}\}$? Its EM wave moves in a direction opposite  
to that in the original solution for electric currents. If interpreted as a 
left-handed way of referring to a wave moving in the {\it same} direction, it is 
the dual of the left-handed vector set \{{\bf E}, -{\bf B}\}. It is actually the 
same physical solution, and it also satisfies the same duality transformation 
(\ref{*Transf}).

\section{Space-time symmetries}\label{symmetries}

All terms of a Maxwell equation have the same symmetries under the space-time 
transformations of parity ${\mathscr P}$: ${\bf r} \rightarrow -{\bf r}$, and time 
reversal ${\mathscr T}$: $t \rightarrow -t$, [\onlinecite{Jackson99-2}]. 
Each term can be 
separated into four parts of different space-time symmetries $(\pm, \pm)$. A classical 
electric charge is a point charge without any spacetime extension. It therefore has 
the {\it intrinsic} symmetries of (even, even) $= (+,+)$ under ${\mathscr P}$ and 
${\mathscr T}$, respectively. These intrinsic symmetries are those of the vacuum. 
Thus the space-time symmetries of $\rho$ are purely external or extrinsic. 
Consider, for example, that part of $\rho$ with symmetries $(+,+)$. Its 
associated {\bf J} is $(-,-)$ in space-time symmetries, its {\bf E} is $(-,+)$, its 
{\bf B} is $(+,-)$, its $_*\rho$ is $(-,-)$ and its $_*${\bf J} is $(+,+)$. 

This symmetry analysis is particularly useful in visualizing the structure of the field 
tensor and field equations. On going from the border vector {\bf E} of the field tensor 
$F$ to its core vector {\bf B}, a time index is changed into a space index. Hence both 
space-time symmetries of {\bf B} must be opposite in sign to those of {\bf E} if one is
dealing with fields of definite symmetries. This explains why in Eq. (\ref{inhomoEq-rt}), 
a spatial current {\bf J} can be generated by both the time variation of the border 
vector {\bf E} and the spatial variation of the core vector {\bf B}. Conversely, each EM 
field can have contributions from both $J$ and $_*J$. 

The {\bf B} field of symmetries $(+,-)$ in the above example can have a contribution 
from the magnetostatic field of a magnetic charge at rest at the origin. The 
external part of such a field has the functional form $\hat{\bf r}/r^2$, where 
$|\hat{\bf r}| = 1$, and the space-time symmetries $(-,+)$. The intrinsic symmetries 
of the magnetic charge must then be $(-,-)$. Hence a magnetic charge cannot be a 
classical point charge with no internal spacetime structure. It is at best 
semi-classical and pointlike.

From the viewpoint of internal symmetries, two other such semi-classical and pointlike 
charges of symmetries  $(-,+)$ and $(+,-)$ are admissible. They satisfy a set of 
gvMaxwell equations that are the opposite-parity partners of the usual set, namely
\begin{eqnarray}
s_GJ'_\mu &=& \partial_\nu F'_{\mu\nu}(x), \nonumber \\
s_G\,_*J'_\mu &=& \partial_\nu \,_*F'_{\mu\nu}(x),
\label{inhomoEq3}
\end{eqnarray} 
where $J'_\mu,\,_*J'_\mu$ are the opposite-parity partners of $J_\mu,\,_*J_\mu$, 
respectively. This completes our derivative of both new and old sets of gvMaxwell 
equations.

The two new charges are similar to magnetic charges in that they have intrinsic symmetries 
different from those of the vacuum. All three charges necessarily have internal structures. 
For a more complete description of the structure and other aspects of magnetic charges, 
see the excellent review by Goldhaber and Trower \cite{Goldhaber90}.

The new set of gvMaxwell equations is mathematically disconnected from the usual set. 
In the quantum theory, their new photons have positive intrinsic parity, opposite in sign
to the negative intrinsic parity of our own visible photons. Since the new photons can be 
emitted or absorbed only by the new opposite-parity charges, both new photons and new 
charges exist only in the parallel universe of dark matter. To inhabitants of this dark 
universe, however, it is our own universe that appears dark.

\end{document}